      [1999/12/01 v1.4c Il Nuovo Cimento]
\documentclass{cimento}

\usepackage{graphicx}  
\title{Hyperon Polarization, Transversity and LHC Physics}
\author{Gary R. Goldstein~\from{ins:x}\ETC,
        \atque
Simonetta Liuti~\from{ins:evil}\thanks{This work is supported in part by the U.S. Department
of Energy grants DE-FG02-01ER4120 (S.L.),  DE-FG02-92ER40702  (G.R.G.).}}
\instlist{\inst{ins:x} Department of Physics and Astronomy, Tufts University - Medford,
       MA 02155 USA.
  \inst{ins:evil} Department of Physics, University of Virginia, Charlottesville, VA 22901 USA.}
\PACSes{PACs 13.60.Hb, 13.40.Gp, 24.85.+p}
\begin{document}

\maketitle

\begin{abstract}
Longstanding puzzles in spin physics can be confronted at the high energies of the LHC. Will large s-quark and c-quark polarization be observed through heavy hyperon production? Top quarks are expected to have significant polarization - single spin asymmetries. How can such observations sort out possible QCD mechanisms? Leptoproduction provides a means to access those mechanisms. TMDs, GPDs and Fracture Functions all provide different model approaches to the polarization phenomena. Their implications will be described.
\end{abstract}

\section{Introduction}
There is a long-standing puzzle in hadronic spin physics, the inclusive production of highly polarized strange hyperons. The expectation from Perturbative QCD based on the Kane, Pumplin, Repko (KPR) calculation~\cite{KPR} was that such polarization should be small as governed by the quantity $\alpha_s(\hat{s}) m_q/\sqrt{\hat{s}}$, where $m_q$ is the effective mass of the strange quark (or the hyperon) that normalizes the helicity flip vertices, and $\sqrt{\hat{s}}$ a characteristic energy of the hard production process for the quark. To explain why measurements did not agree with the expectation, soft processes or semi-classical mechanisms of various kinds have been proposed over the years. To this end, Dharmaratna and Goldstein (DG)~\cite{Dharma} calculated all the 4th order hard QCD amplitudes that enter in hadron+hadron collisions via parton+parton producing s-quark pairs or heavier flavors. They found that sizable polarizations, at the several percent level, were produced through interference between these amplitudes and tree level terms. The kinematic dependence of these results favored small $p_T$. Negative $x_F$ was linked to negative polarization. The gluon fusion mechanism produced the larger magnitude of polarization and is antisymmetric around the 90$^\circ$ production in the g+g center of mass. If this hard process is the source of measured hyperon polarization, that polarization must be enhanced through soft process  hadronization. Furthermore, the $x_F$ must be boosted to the forward region. It was posited that this was accomplished through recombination of the polarized quark with the diquark remnant of the beam proton. A simple ansatz for that soft process used the analogy with the ``Thomas precession" mechanism~\cite{Thomas} to boost the $x_F$(quark) to $x_F$(hadron) while enhancing the scale of the polarization by $\approx 2\pi$. This DG scheme explained the systematic behavior of $\Lambda$ and $\Sigma$ polarization. But it required this ``recombination" to enhance the s-quark polarization while realigning the kinematics. Taken at face value it predicted~\cite{GGLambda_c} the behavior of the $\Lambda_c$ data from Fermilab experiment E791~\cite{E791}. The latter show increasing negative polarization for $p_T\approx 2$ GeV. This striking result awaits confirmation from higher energy collisions at the LHC. The attractive features of the DG ``hybrid model" suggest  a more rigorous QCD based formulation. These will be elucidated below. First we consider the production of top quarks.

\section{Top quark polarization}

\begin{figure}
\centerline{
\includegraphics[width=8.cm]{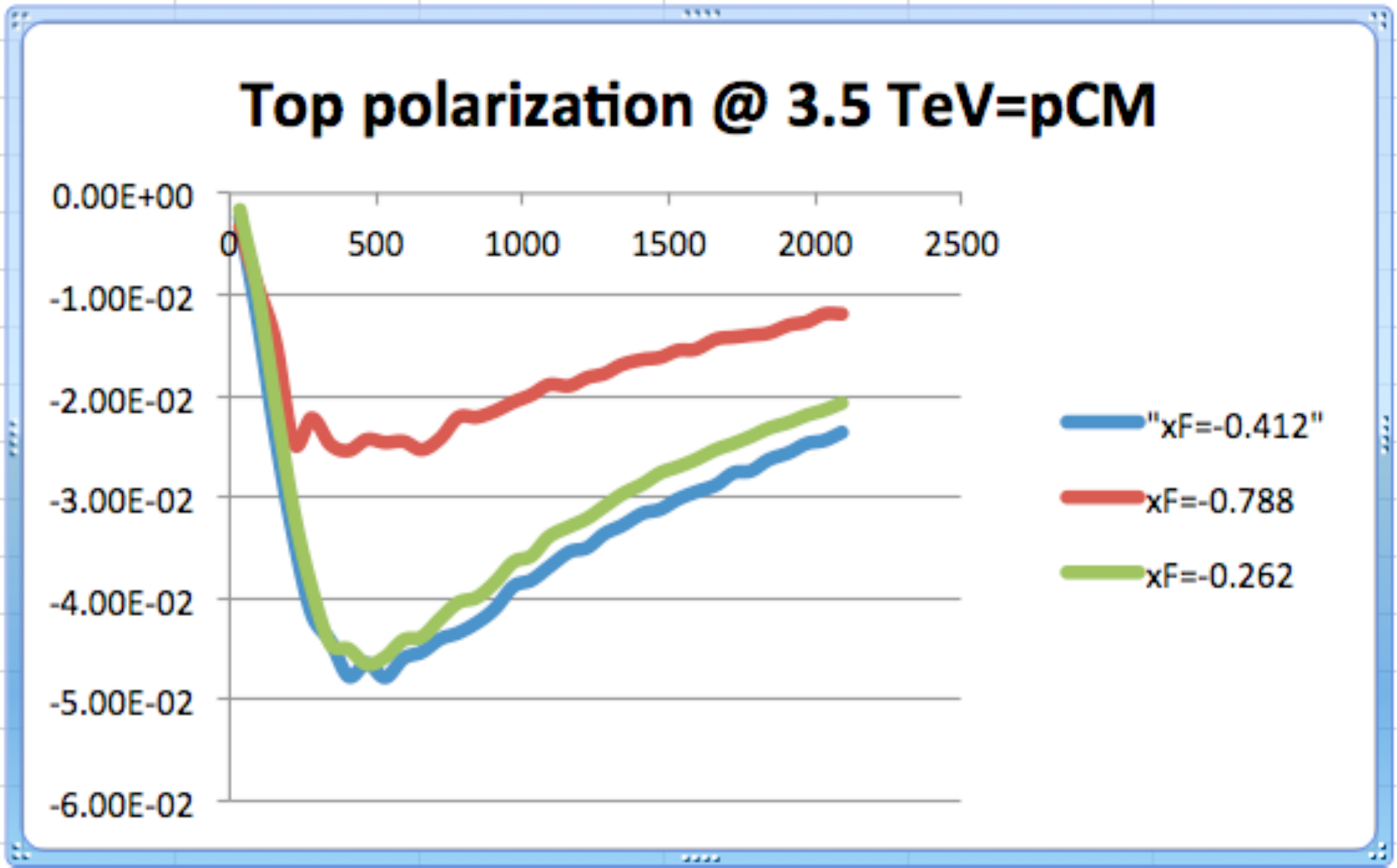}}
\caption{
TOP QUARK POLARIZATION prediction for LHC energy at several Feynman x values - preliminary calculation.}
\label{t-polzn}
\end{figure}

The hard processes of gluon fusion and light quark pair annihilation as mechanisms to produce a heavy quark polarization were shown to reach the few percent level in the calculations of DG. These calculations were carried out to one loop order in QCD as necessary for non-zero polarization. For the c-quark and b-quark the polarization has to be carried through the hadronization, and hence soft processes will intervene to produce the polarized hadron. For {\bf top quarks}, however, hadronization does not occur - the top decays weakly, but faster than hadronization times, primarily into $W^+ + b$~\cite{dalitz}. The orientation of the decay plane relative to the production plane is a measure of the polarization. Preliminary calculations by Goldstein and Liuti are shown in fig.~\ref{t-polzn}.  
This figure shows that, remarkably, there can be up to 5\% polarization over a very broad range of transverse momenta for LHC kinematics. Qualitatively, the large top mass enhances the polarization overall. The falloff with the energy production scale is compensated by the large gluon production probability, especially at small values of $x_{Bj}$. There is also a notable scaling up of the range of transverse momenta as the phase space becomes very large compared to fixed target kinematics. The net effect is that the gluon distribution functions contribute most to the smaller Feynman x region. A measurement of this quantity will provide a good test of the PQCD hard production process, independent of soft rescattering effects.

\section{Leptoproduction mechanisms}

Several directions are being pursued to explore the hyperon and heavy flavor baryon polarization phenomena as applied to  LHC energies, where the $g + g$
mechanism is dominant, especially at small Bjorken $x$ for both protons. To begin with, the leptoproduction framework is used in preparation for considering purely hadronic production. One approach is through the 
``fracture function" picture of Trentadue and Veneziano~\cite{TV}, as extended to include angular momentum dependences by Sivers~\cite{sivers1}. The fracture functions are joint probabilities that relate the beam direction hard scattered quark distribution from the target nucleon (a function of $x_{Bj}$) to the baryon fragmentation of the target (a function of $z$). For a single spin asymmetry (SSA) there must be an interference, which requires some process like a final state interaction. Another approach is from the GPD perspective. The Compton Form Factors derived from GPDs have phases that will provide such interferences. We will illustrate both approaches.

\subsection{Fracture Function approach}

The Fracture Functions (FF) provide a QCD-based perspective on target fragmentation. In leptoproduction of hadrons, the FF represents the joint probability for the emission from the hadronic target of a quark, antiquark or gluon at some $x = Q^2/2M\nu$ along with the fragmentation of the target into a hadron of fractional momentum $z=E_{hadron}/E(1-x_{Bj})$ relative to the target energy in the virtual photon +  target CM frame. This definition of $z$ establishes that for large $z$ the hadron is tracking the fragment of the target, {\it e.g.} the diquark or spectator. Formally, for a quark-nucleon FF,
\begin{eqnarray}
\mathcal{F}^{\lambda_q}_{\Lambda_N; \Lambda_{\Lambda}^\prime,\Lambda_{\Lambda}}(x, k_T, z, p_T,Q^2)&=&\sum_{\Lambda_X} \int \frac{d^3P_X}{(2\pi)^3 2E_X} \int \frac{d^4\xi}{(2\pi)^4} e^{i k\cdot \xi} 
\nonumber \\
\times  \langle P,\Lambda_N\mid {\bar \psi}^{\lambda_q}(\xi) \mid P_{h},\Lambda_{\Lambda}^\prime ; X \rangle  & \times& 
 \langle P_{h},\Lambda_{\Lambda} ; X \mid \psi^{\lambda_q} (0) \mid P,\Lambda_N \rangle.
\label{FF}
\end{eqnarray}
where transverse momenta $k_T, p_T$ are for the quark and the outgoing hadron, respectively. Since we are interested in the unpolarized cross section or the transverse $\Lambda$ polarization, we keep the definition here in terms of diagonal helicities for the quark fields and the target nucleon, while allowing for change of helicity for the outgoing hadron. It should be noted that there can be transverse polarization providing the $\Lambda$ has non-zero $p_T$ relative to the $\gamma^*$ direction. This is possible, whether a current fragment is observed or not, because there is a hadronic plane established. This can be a leading twist result (although outside the scope of the thorough formulation in ref.~\cite{ABK}), by analogy with the ref.~\cite{BHS} for Semi-Inclusive Deep Inelastic Scattering (SIDIS), in which a soft gluon exchange provides the necessary phase. 

We evaluate the behavior of FFs in describing leptoproduction of polarized hyperons within the spectator model - the diquark spectator model~\cite{GGL2}. This model has provided a successful beginning in qualitatively describing a large array of phenomena. We first consider a FF at the tree level for the process $\gamma^* + N \rightarrow {\bar s} + \Lambda + X$ where the ${\bar s}$-quark hadronizes and $X$ contains unmeasured, inclusive hadrons. 
In the spectator model for the nucleon to a light quark transition, the spectator is the (uu) or (ud) diquark. We take a scalar diquark to begin with, (ud+du). Then we can write a form in which the struck u-quark (with $x$ and $k_T$) leaves the diquark (with $(1-x)$ and $-k_T$) to fragment into the $\Lambda$ (with $z$ and $p_T$).  

\begin{figure}
\centerline{
\includegraphics[width=5.cm]{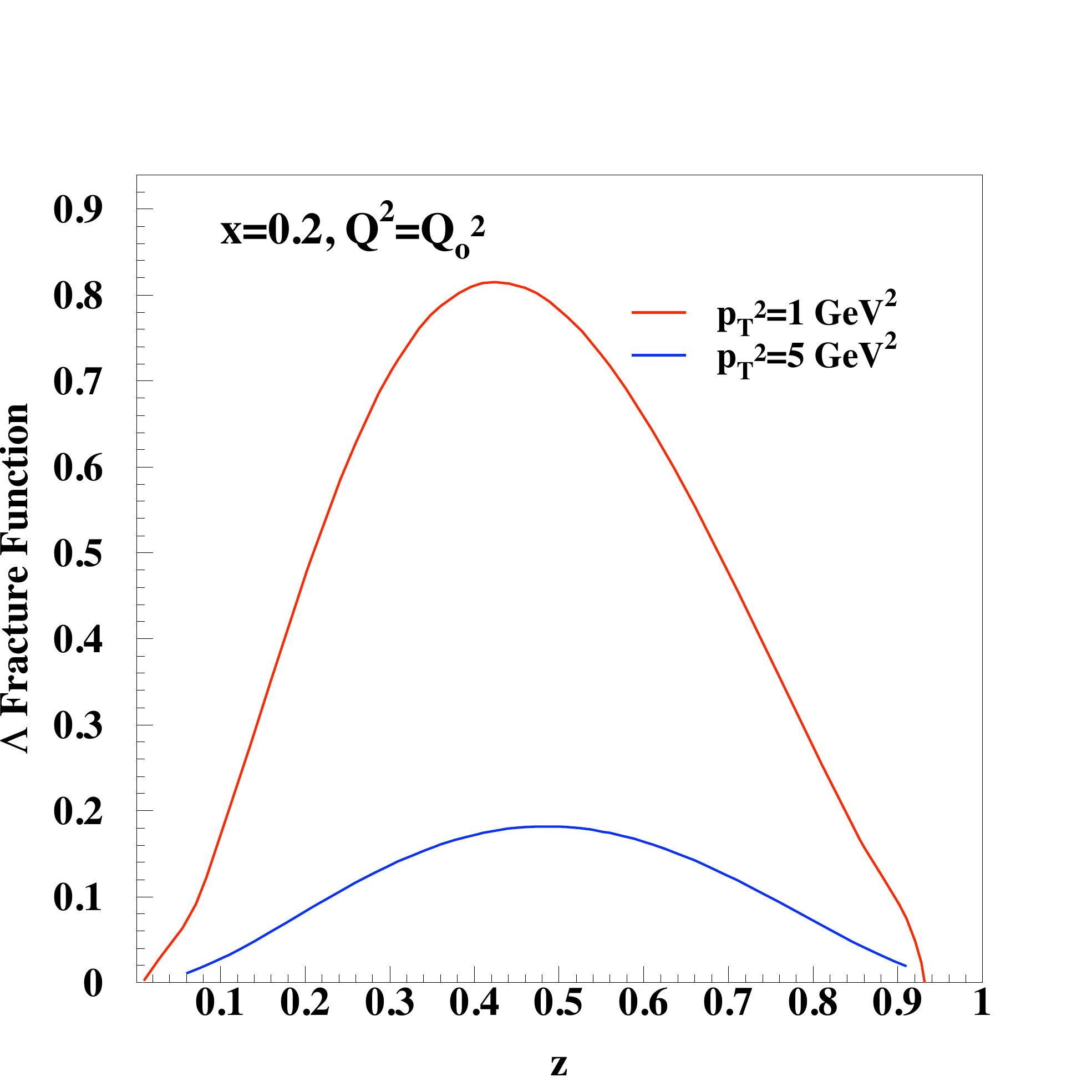}
\hspace{0.5cm}
\includegraphics[width=7.cm]{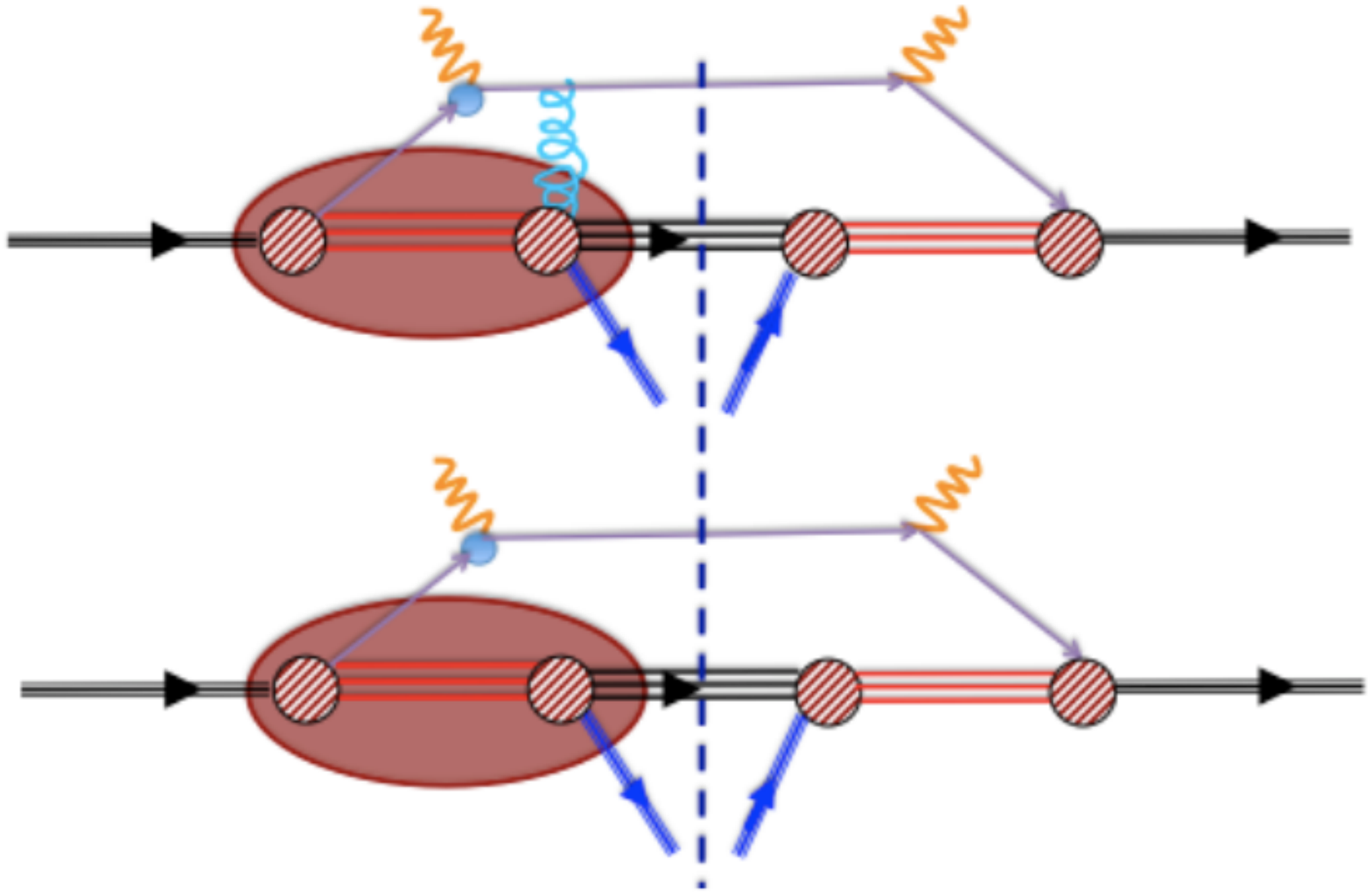}}
\caption{(LEFT) Tree level Fracture Function for $\Lambda$ production at $x=0.2$ for two values of $p_T$. (RIGHT) Diagram for polarized $\Lambda$ fracture function (upper graph); tree level contribution to unpolarized $\Lambda$ fracture function (lower graph). }
\label{fig4_FF}
\end{figure}

The flavor labels have been ignored. Obviously, for polarized $\Lambda$ production, those labels are crucial. It is also important to note that gauge links have to be carefully implemented for consistency. The latter can be interpreted as final state interactions, via the mechanism of ref.~\cite{BHS}. In fact, to obtain polarization in this picture requires some means to generate a phase difference between interfering helicity flip and non-flip amplitudes. As Sivers has noted~\cite{sivers1}, there has to be a means to generate a ``Boer-Mulders" FF or a ``Polarizing"  
FF in order to have an unpolarized target yield a polarized $\Lambda$. The mechanism we will consider is illustrated by having an additional gluon exchange as in the upper right diagram of 
fig.~\ref{fig4_FF} 
connecting the outgoing parton to the spectator or the fragmenting inclusive state $X$.

At tree level the calculation of the cross section is straightforward. 
To approximate all of those states and in keeping with the spectator approach, we take X to be a single mass $m_s$ anti-s quark. Then there is the condition that $(P-k-P_\Lambda)^2 = m_s^2$. Given this condition we have the diquark momentum $(P-k)^2$ and the quark momentum $k^2$ constrained to be
\begin{eqnarray}
(P-k)^2&=&\frac{m_s^2}{1-z} + \frac{M_\Lambda^2}{z} + \frac{1-z}{z} \left( {\vec P}_{\Lambda T} - \frac{z}{1-z}{\vec P}_{XT} \right)^2 \nonumber \\
k^2&=&xM^2-\frac{{\vec k}_T^2}{(1-x)} - \frac{x}{(1-x)} (P-k)^2, \;\;\;\;  
k^+ = x P^+
\label{constraints}
\end{eqnarray} 
The struck quark with momentum $(k+q)$ leads to $\delta((k+q)^2)\approx \frac{x_{Bj}}{Q^2} \delta(x - x_{Bj})$. 

The unpolarized fracture function is evaluated in this model by summing over the unobservable helicities and integrating over 4-momentum $k$. 
If the struck quark is not ``observed", {\it i.e.} its fragmentation is not measured, the $\vec{k}_T$ can be set to 0 without loss of generality. Then we have 
the resulting Fracture Function $\mathcal{F}$, shown for two values of $p_T$ in fig.~\ref{fig4_FF}, left.
The corresponding diagram is the lower right diagram of fig.~\ref{fig4_FF}.

For the SSA, amplitudes with final state interactions can be evaluated in analogy with TMDs (see refs.~\cite{GGS} \& \cite{BCR} calculations). There are two ways the final state interaction can occur at order $\alpha_s$. The color non-singlet diquark can absorb the soft gluon from the struck quark or the fragment remnant from the $\Lambda$ production with anti-s-quark quantum numbers can absorb the soft gluon. This is shown in the upper right fig.~\ref{fig4_FF} as the gluon being absorbed by the diquark to form the $\Lambda$ and X.
The first possibility is completely analogous to the calculation of transversity-odd TMDs, the Sivers and Boer-Mulders functions, for SIDIS except for the fragmenting of the outgoing diquark. For that case the resulting form of the Polarization for the TMDs~\cite{BHS,GGO} is
\begin{equation}
P_y=C_F \alpha_s(\mu^2) \frac{(xM+m)k_x}{\left[(xM+m)^2+{\vec k}_\perp^2\right]}\frac{\Lambda({\vec k}_\perp^2)}{{\vec k}_\perp^2} ln\left(\frac{\Lambda({\vec k}_\perp^2)}{\Lambda(0)}\right)
\label{tmd_pol}
\end{equation}
with
\begin{equation}
\Lambda({\vec k}_\perp^2)= {\vec k}_\perp^2 + x(1-x)\left(-M^2+\frac{m^2}{x}+\frac{\lambda^2}{1-x}\right)
\label{tmd_kallen}
\end{equation}
Note that this corresponds to the product of an imaginary part of the amplitude from the one loop diagram times the real part of the tree level amplitude all divided by the tree level amplitude squared. The fracture function of fig.~\ref{fig4_FF} has kinematics that replace the designations of Eq.~\ref{tmd_pol} with the apprpropriate denominator - 
a function of both the quark and the $\Lambda$ transverse momenta and their corresponding longitudinal momentum fractions $x,\, z$.

Once this picture is completed for the leptoproduction, the hadronic process, $p+p\rightarrow \Lambda +X$ has to be tackled. For the right kinematics the virtual photons in right fig.~\ref{fig4_FF} can be replaced by gluons or antiquarks originating from the other proton. Results will be presented elsewhere.

\subsection{GPD approach}

In the GPD description of $\gamma^*+N \rightarrow {\bar s} + \Lambda$, the process amplitudes are linear combinations of the complex Compton Form Factors (CFFs). The CFFs are obtained from the integration of the real GPDs over the unobserved quark or gluon momentum fraction of the target momentum. For
${\mathcal H}_{N \rightarrow Y}$, a generic example of a CFF for one of the flavor changing GPDs is
\begin{eqnarray}
{\mathcal H}_{q,g} = \int dX H_{q,g}(X,\zeta,t) \left[ \frac{1}{X-\zeta - i \epsilon} + \frac{1}{X - i \epsilon} \right]
\label{CFF}
\end{eqnarray}

GPDs  are parametrized as functions of $(X,\zeta,t,Q^2)$ (for notation and parameterization of the full set of spin dependent GPDs see ref.~\cite{GGL}). Again, we are interested in the $p+p$ process, so the virtual photon will have to be replaced by an antiquark or a gluon. Then the variables are related as 
$x_1 \equiv X$, the quark, gluon or antiquark fraction of the initial proton
momentum, $x_F=1-\zeta >0$,  the hyperon's momentum fraction, $p_T^2 = \Delta_\perp^2$ which is related to $t$. The scale of the process is $Q^2$. 
  
When the protons interact through having virtual photons replaced by antiquarks or gluons, then the general form of the cross section will be 
\begin{eqnarray}
& \displaystyle\int & d x_2 \; \left[ {\mathcal H}^*_{N \rightarrow Y}  {\mathcal H}_{N \rightarrow Y}\right](\zeta(x_F),t(p_T^2),Q^2)   \nonumber \\ 
 & \times & f(x_2,Q^2) \hat{\sigma}_{1 2 \rightarrow s X}  (x_2, x_F^s, p_T)  
\end{eqnarray}
where $N\rightarrow Y$ represents the nucleon to hyperon transition. Flavor off-diagonal quark correlators at the amplitude level are

\begin{eqnarray}
\label{correlator}
&& \int d z^- e ^{i \, q^+ z-} \langle \Lambda \mid \bar{\psi}^a (z^-) \gamma^+ \psi^b(0) \mid P \rangle =  
\nonumber \\
&& \overline{U}(\Lambda) \left[ H_{N\rightarrow\Lambda}(X,\zeta,t) \, \gamma^+ + E_{N\rightarrow\Lambda}(X,\zeta,t) \, \frac{-i \sigma^{+,\lambda}}{2M} \Delta_\lambda \right] U(P),  \nonumber \\
\end{eqnarray}
where $(a,b)$ are appropriate flavor labels, {\it e.g.} $a=u$,$b=s$, that are to be contracted with the flavor changing PQCD hard subprocess.
The hard subprocess involves helicity conserving interactions for zero mass quarks. The only contributions to the overall convoluted helicity amplitudes will be those for which the Extended GPD ($p\rightarrow u : s \rightarrow \Lambda$) has the form 
$A_{\Lambda_{\Lambda}, \lambda_s=+; \Lambda_p,\lambda_u=+}$ or $A_{\Lambda_{\Lambda}, \lambda_s=-; \Lambda_p,\lambda_u=-}$. We evaluate these  off-diagonal GPDs using the parameterization of ref.~\cite{GGL2,GGL,AGL} with SU(3)$_{flavor}$ for the Lambda-diquark-quark vertex. Once the off-diagonal CFFs have been determined, the resulting complex, helicity dependent amplitude structures are  folded into the $p+p \rightarrow \Lambda + X$ helicity amplitudes to obtain cross sections and polarizations. This will be presented elsewhere. 

\acknowledgments
We are grateful to the organizers of Transversity 2011 for producing an interesting and productive workshop in a lovely setting.

\end{document}